# A Neighbour Search Code for Galaxies-Preliminary Results

René A. Ortega-Minakata[1], Juan P. Torres-Papaqui1, Heinz Andernach[1]

*[1] Departamento de Astronomía, Universidad de Guanajuato*

*{rene, papaqui, heinz}@astro.ugto.mx*

## Abstract

*We present preliminary results of appl-ying a neighbour search code to a large sample of galaxies drawn from the Sloan Digital Sky Survey (SDSS). We draw our sample of target galaxies from the spectroscopic catalogue of the SDSS, which has redshift measurements for all its ~ 8 x 105 galaxies. We count the neigh-bours from the photometric catalogue of the SDSS, which is a flux-limited catalogue that has a photometric redshift estimation for all its ~ 2 x 108 galaxies.*

*The code counts the number (N) of galaxies within a given 3D-distance from the target galaxy, using a database manager and a simple decision tree. The 3D-distance is com-posed of a 2D projected distance on the sky and a radial component, which is the difference between the spectroscopic target redshift and the photometric redshift of the "neighbour" galaxies. As we keep the search volume cons-tant throughout the sample, N is a proxy for the number density of galaxies around the target galaxies.*

*We also present an analysis of the pre-liminary results, showing the relation between N, the inferred morphology and the dominant activity type of the target galaxies, as well as the relation between N and the star formation history of the target galaxies.*

## 1. Introduction

In our project, we wish to quantify the number density of galaxies in three-dimensional space to investigate whether some statistical properties of galaxies, specifically their type of dominant activity, morphology and star formation history, are related to this density. This requires a neighbour search algorithm applied to a large sample of galaxies, which we call the target sample, and a very large sample of galaxies to search the *neighbours* from, referred to as the *neighbours* sample.

The density of the environment of a galaxy is difficult to define, as the best measurement of environment depends on the particular goals of each study (for a detailed discussion, see [1]). It is usually defined as the number of galaxies around a particular galaxy or in a given area of the sky. Different particular definitions yield results with small variations (for examples see section 5 in [2]), so an adequate definition of environment is crucial to achieve our goal. We have chosen a parameter represented as $N_R^{-19}$, based on a similar measurement by [3], which is the number of galaxies with absolute r-band magnitude $M(r)$ less than or equal to -19 within a physical projected search radius R of a target galaxy.

## 2. The Algorithm

Basically, our algorithm counts the number (N) of galaxies within a certain search radius in three-dimensional space of a target galaxy. The relevant information of the galaxies in both the target and the neighbours samples is their position in three-dimensional space, which is given in spherical coordinates by the angular coordinates in the plane of the sky (RA and Dec), and a coordinate along the radial direction, the redshift z.

The search radius in three-dimensional space can be decomposed into a component in the plane of the sky, equivalent to a physical projected search radius, R, and a radial component, equivalent to a difference in z. This decomposition yields a truncated conical search region rather than a spherical one which is adequate for the purpose of this code. It is also worth noting that we choose the physical size R as constant for all the target galaxies, so when expressed as an angular distance $\theta R$ in the plane of the sky, $\theta_R$ decreases with the redshift of the target galaxy. This also implies that the search volume remains constant, so the number of galaxies within any such *volume* is directly proportional to the density of galaxies in that region.

We chose to draw our input samples from the Sloan Digital Sky Survey Data Release 7 (SDSS-DR7) [4]:

• The target sample consists of $\sim 8 \times 10^5$ galaxies from the spectroscopic catalogue. All galaxies in this sample have measured redshifts.

• The neighbours sample comprises $\sim 2 \times 10^8$ galaxies from the photometric catalogue. This is reduced after applying relevant selection criteria, mainly the aforementioned limit in the absolute magnitude in the r band $M(r)$ less than or equal to -19, the sole application of which yields $\sim 1.14 \times 10^8$ galaxies. It is important to note that we only have estimated photometric redshifts for all these galaxies. For a reference set of galaxies with spectroscopic redshifts, the rms error of the redshift estimation is 0.025 [4].

The basic algorithm consists of two nested loops. The outer loop sweeps through all the galaxies in the target sample, reads the relevant data for each of them and calculates the range in redshift and projected angular search radius $\theta R$ in which neighbours need to be searched. The inner loop sweeps through the galaxies in the neighbours sample, reading the relevant data for each of them; for those within the current target's range in redshift, it calculates the projected angular distance from the target; it then counts all galaxies with projected angular distance within $\theta_R$. Finally, the outer loop writes the final count for each target galaxy.

We coded a working version of this basic algorithm in the Perl Data *Language, PDL* (see pdl.perl.org). Since this basic algorithm would take a very long time to run using the full samples described above, we needed to find optimization strategies to reduce the run time of the code.

Some of these strategies are:

• Parallelization: both loop structures

are suitable for parallelization, since results of each comparison and also of a previous target are independent of any previous comparison or target.

• Relevant region search: we do not need to test all objects in the neighbours sample but only those within a relevant region near the target; we need to skip to the next target after searching this region.

So far we have only pursued a relevant-region search strategy, taking advantage of Perl modules that connect with database managers; such as PostgreSQL (see www.postgresql.org). We coded a version of the algorithm that queries the neighbours sample, pre-loaded into a PostgreSQL database, for galaxies within the relevant redshift range and inside a "square" projected into the plane of the sky with dimension just large enough to contain a circle of radius R around the target galaxy.

This last version of the code was tested against the basic algorithm with no optimization strategy, finding that the optimized code was ~ 65 times faster. A parallelization strategy has yet to be implemented, but a preliminary analysis that makes use of the results of the test described in the next section suggests that by using a process manager to run different regions of the sky simultaneously would be enough to run the code for the full samples mentioned above in a reasonable time (see also the last paragraph of the next section).

## 3. Testing the Algorithm.

Using the optimized version, we tested the algorithm for a region of sky between +20° and +30° of Declination and 170° and 180° of Right Ascension. This region of the sky comprises 10,860 galaxies from the SDSS-DR7 spectroscopic catalogue, making the target sample, and 814,704 galaxies from the SDSS-DR7 photometric catalogue with M(r) less or equal than -19, making the neighbours sample. We present here basic statistics regarding this test.

We ran the test using a physical projected search radius R = 1.5 Mpc, a radial velocity (c·z) range of plus or minus 2500 km/s from the target redshift, and a value of the Hubble constant $H_0 = 75$ km·s-1·Mpc-1, which measures the rate of the expansion of the Universe. The median value of the number of neighbouring galaxies was $N_R\text{-}^{19} = 9$. In Figures 1 and 2 we present, respectively, the histogram of $N_R\text{-}^{19}$ and the distribution of $N_R\text{-}^{19}$ with the redshift z of the target galaxies. For comparison, we present the histograms of the redshift z of both the neighbours and target galaxies in Figure 3. Figure 4 displays the distribution on the plane of the sky of all target galaxies, coloured according to their value of $N_R\text{-}^{19}$.

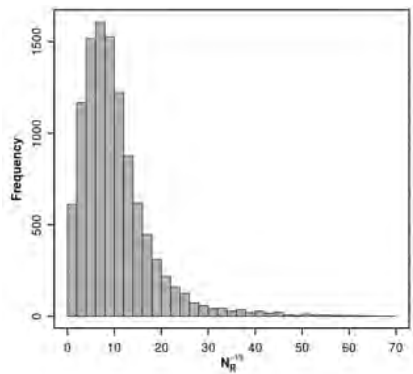

**Figure 1. Histogram of $N_R\text{-}^{19}$.**

While making this test, we checked for possible inconsistencies that would indicate a failure of the algorithm, such as the mean photometric redshift of the galaxies returned by the SQL query being very different from the target galaxy redshift, or the number of neighbours being higher than the number of galaxies returned by the SQL query. We found none of these inconsistencies, suggesting the code works appropriately.

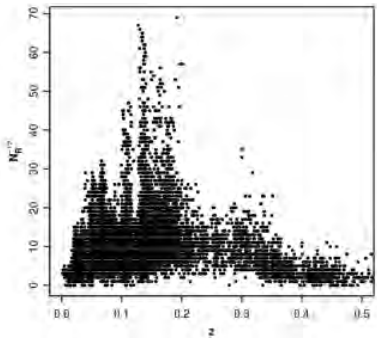

**Figure 2. Distribution of NR-19 with the redshift z of the target galaxies, out to z = 0.5 .**

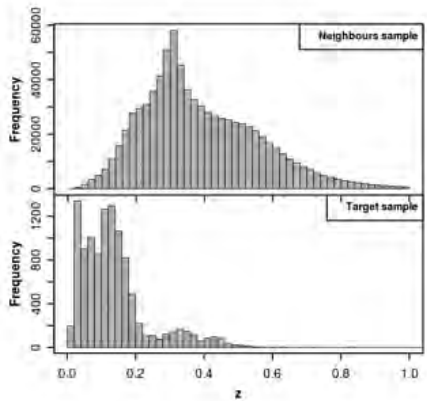

**Figure 3. Histogram of redshift z of the neighbours sample (upper panel) and of the target sample (lower panel).**

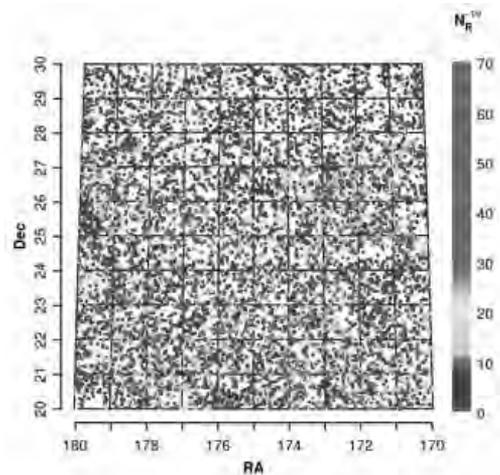

**Figure 4. Aitoff projection of the distribution on the plane of the sky of all target galaxies, coloured using rainbow colours according to their value of $N_R^{-19}$: $N_R^{-19}$ = 0 is represented with magenta, running smoothly to $N_R^{-19}$ greater or equal than 30, represented with red.**

Proper comparison with similar measurements of galaxy environment in the literature is work in progress. Still, we can already conclude that the code as it is fulfills its purpose of calculating the number of galaxies within a given search radius. Parallelization of the code, which is feasible given its structure, can be achieved either by direct parallelization or by running the code for different regions of the sky simultaneously, and will allow us to run it on the above mentioned samples of the order of 800,000 target galaxies and 114 million galaxies in the neighbours sample, completing our original goal. Such a run would take approximately 24 hours using 205 simultaneous processors.

## 4. Testing our relations between activity type, morphology, star formation history and environment using our measurement of environment

Using the results of the test described in the previous section, we explored the relations between dominant activity type, morphology, star formation history (SFH) and environment of galaxies that we had found previously (see [5]), but now using our measurement $N_R^{-19}$. In general, galaxies may show emission-line activity that can be dominated by either an active galactic nucleus (AGN) or star formation (SFG), or the galaxy may be a composite "transition object" (TO), while the morphology of a galaxy can generally be defined as "early type" (elliptical and lenticular galaxies) or "late type" (spiral and irregular galaxies). Since our target sample is too large for a visual inspection of the morphology, we use a morphological inference derived from photometric parameters of the galaxies. For detailed descriptions on how these characteristics of galaxies are defined and used here, see [5], [6], [7], [8] and references therein.

For this test, we have restricted ourselves to galaxies within the range 0.025 to 0.250 in z, in order to minimize aperture biases (lower constraint) and to avoid the need of a correction to the luminosity function due to the Malmquist bias (upper constraint). We considered seven bins of NR-19, as defined in Table 1. Note that we do not use NR-19 = 0, since this indicates a failure in the algorithm for that particular galaxy, because NR-19 does include the target galaxy. These failures are mostly due to larger uncertainties in the photometric redshifts of these galaxies. There are 8831 galaxies within the aforementioned redshift range, of which very few (13) have NR-19 = 0. This is also a good indicator that the code is working properly. The total number of galaxies used for our test is therefore 8818.

**Table 1. Activity type distribution for different neighbourhood densities.**

| Bin no. | $N_R^{-19}$ | Number of Galaxies |||||
|---|---|---|---|---|---|---|
| | | Total | SFG (%) | AGN (%) | TO (%) | Passive (%) |
| 1 | 1 to 3 | 461 | 183 (39.7) | 69 (15.0) | 70 (15.2) | 6 (1.3) |
| 2 | 4 to 6 | 1536 | 536 (34.9) | 275 (17.9) | 225 (14.6) | 29 (1.9) |
| 3 | 7 to 9 | 2045 | 653 (31.9) | 407 (19.9) | 303 (14.8) | 48 (2.3) |
| 4 | 10 to 14 | 2537 | 692 (27.3) | 551 (21.7) | 334 (13.2) | 80 (3.6) |
| 5 | 15 to 19 | 1171 | 260 (22.2) | 253 (21.6) | 164 (14.0) | 43 (3.7) |
| 6 | 20 to 29 | 745 | 135 (18.1) | 163 (21.9) | 80 (19.7) | 38 (5.1) |
| 7 | 30 or more | 323 | 36 (11.1) | 72 (22.3) | 30 (9.3) | 15 (4.6) |

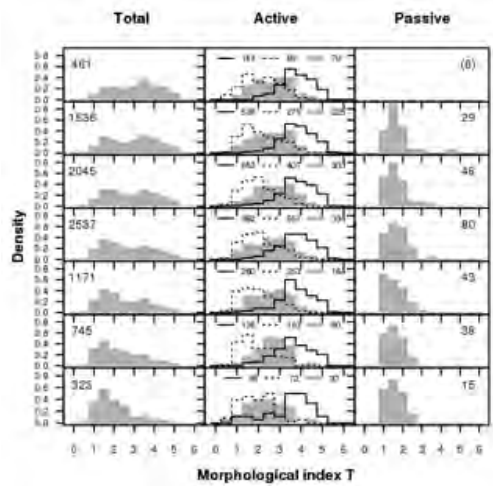

**Figure 5. Morphological distribution of galaxies within the different bins of NR-19. Bin 1 is the top row, while Bin 7 is the bottom row; for the total number of galaxies in each bin (left column), separated by activity type (central column), and for passive galaxies (right column). In the central column, the solid line histograms corresponds to SFG, the dotted line histograms to AGN and the gray-shaded histograms to TO. The sizes of the samples are indicated. Areas below all histograms are equal to unity. The top right panel is omitted because of the very small amount of passive galaxies in Bin 1 (only 6).**

The results of this test can be seen in Figures 5 and 6. In Figure 5 we show the morphological distribution of galaxies within the different bins of $N_R^{-19}$, in general and separated by dominant activity type. The left column clearly shows the morphology-local density relation [9], as the fraction of early-type galaxies clearly increases from lower density (top) to higher density environments (bottom). The central column shows the active galaxies by type: AGN-dominated (AGN), Star Formation-dominated galaxies (SFG) and Transition Objects (TO). The morphological separation between these types of dominant activity, seen previously using ACO clusters [10] and isolated galaxies [11], holds independently of $N_R^{-19}$, which means that there is a clear relation between the inferred morphology and the dominant type of activity of the galaxies.

In combination with Table 1, we can see how the fraction of SFGs decreases sharply with increasing

$N_R^{-19}$, which can be a consequence of the morphology-density relation combined with the morphology-activity relation we have confirmed. In the right column of Figure 5 we can see that passive galaxies are always of early morphological type, also independently of $N_R^{-19}$, and that their fraction also increases with increasing environmental density.

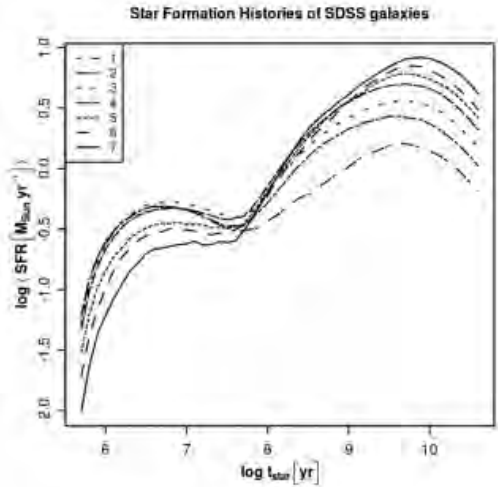

**Figure 6. Median star formation histories (SFHs) of galaxies within the different bins of $N_R^{-19}$. The curves of SFHs are formed by relating the star formation rate (SFR) needed at a given lookback time $t_{star}$ to form the current stellar populations of each galaxy, so a higher value of SFR at a given $t_{star}$ value means that a higher fraction of stars with age tstar is present in that galaxy.**

In Figure 6 we show the median SFHs of galaxies within the different bins of $N_R^{-19}$. A very clear sequence of SFHs can be distinguished, ranging from the denser to the less dense environments: denser environments have higher first peaks of star formation (higher fraction of older stellar populations) and lower recent peaks (lower fraction of younger stellar populations), while less dense environments have lower first peaks and higher recent peaks of star formation. This is also in agreement with previous results using extreme environments (rich ACO clusters and isolated galaxies), confirming that there is an actual sequence of SFHs from low-density environments to high-density environments.

## 4. Conclusions

We have presented preliminary results of applying a neighbour search code to a large sample of galaxies drawn from the SDSS.

The code counts the number (N) of galaxies at a given 3D-distance from the target galaxy, using a database manager and a simple decision tree. The 3D-distance is composed of a 2D projected distance on the sky and a radial component, which is the difference between the spectroscopic target redshift and the photometric redshift of the neighbour galaxies.

We have described how this code works and tested it using a sample of 10,860 target galaxies and 814,704 galaxies in the neighbours sample, drawn from the SDSS-DR7, showing that the code fulfills its goal and can be run on the full sample in reasonable execution time.

We also presented an analysis of the preliminary results, showing the relation between N, the inferred morphology and the dominant activity type of the target galaxies, as well as the relation between N and the star formation history of the target galaxies. Our environment measurement is able to reproduce the morphology-local density relation, and is independent of the morphology-activity relation. We also show a clear sequence of SFH with N, showing that galaxies in denser environments tend to be older than galaxies in less dense environments.